%% file: main.tex
\documentclass[aps,twocolumn,showpacs,superscriptaddress,groupedaddress]{revtex4}  
\usepackage{graphicx}  
\usepackage{dcolumn}   
\usepackage{bm}        
\usepackage{amsmath,amssymb}   
\usepackage{xcolor}   
\usepackage{float}   
\usepackage{lipsum}
\usepackage{subfiles}
\usepackage{ragged2e}

\let\vaccent=\v 
\renewcommand{\v}[1]{\ensuremath{\mathbf{#1}}} 
\renewcommand{\t}[1]{\ensuremath{\text{#1}}} 

\begin{document}

\widetext


\title{\LARGE Navigating at Will on the Water Phase Diagram
}
\author{S.~Pipolo}
\email{silvio.pipolo@univ-lille.fr}
\altaffiliation{Present affiliation: Univ. Lille, CNRS, Centrale Lille, ENSCL, Univ. Artois, UMR 8181 - UCCS - Unit\'e de Catalyse et Chimie du Solide, F-59000 Lille, France}
\affiliation{Sorbonne Universit\'es, UPMC Univ. Paris 06, CNRS UMR 7590, IRD UMR 206, MNHN, IMPMC, F-75005 Paris, France}
\author{M.~Salanne} 
\affiliation{
Sorbonne Universit\'es, UPMC Univ. Paris 06, CNRS, Laboratoire PHENIX, F-75005 Paris, France}
\author{G.~Ferlat} 
\author{S.~Klotz} 
\author{A.M.~Saitta}
\author{F.~Pietrucci}
\email{fabio.pietrucci@upmc.fr}
\affiliation{Sorbonne Universit\'es, UPMC Univ. Paris 06, CNRS UMR 7590, IRD UMR 206, MNHN, IMPMC, F-75005 Paris, France}
\date{\today}

\begin{abstract}
Despite the simplicity of its molecular unit, water is a challenging system because of its uniquely rich polymorphism 
and predicted but yet unconfirmed features. 
Introducing a novel space of generalized coordinates that capture changes in the topology of the interatomic network, 
we are able to systematically track transitions among liquid, amorphous and crystalline forms throughout the whole phase diagram of water, 
including the nucleation of crystals above and below the melting point. 
Our approach, based on molecular dynamics and enhanced sampling / free energy calculation techniques, 
is not specific to water and could be applied to very different structural phase transitions, 
paving the way towards the prediction of kinetic routes connecting polymorphic structures in a range of materials.
\end{abstract}
\maketitle

Computational structure prediction methods~\cite{pickard2011ab,glass2006uspex} have strongly contributed to the rapid increase of new predicted phases of materials with enhanced properties for applications (see, e.g., Ref.~\cite{wilmer2012large}). However, at present, no general approach has been developed for guiding experiments through the pathways connecting stable structures of condensed matter.
Moreover, metastable phases are very often involved in phase
transitions~\cite{schreiber2016real} and sometimes their kinetic stability is very high~\cite{himoto2014yet}. Thus, in order to recover the global minimum structure, one needs to find specific routes, by e.g. acting on pressure or temperature, in a way that is not at all trivial to guess~\cite{radha2015thermodynamic}.
A precise understanding of transition mechanisms and the corresponding kinetics is therefore the key to explain and control the behavior of matter.
The case of water is emblematic because several experiments have disclosed connections between stable and metastable  
phases~\cite{bartels2012ice,palmer2014metastable,mishima1998relationship,klotz2005t}
and recently simulations have highlighted the importance of metastable states in understanding the mechanism of phase transitions and related 
transformations~\cite{russo2014new}.
A classic example is the connection between the crystalline ice stable at ambient pressure (Ice~I), and the low-density amorphous (LDA) and high-density amorphous (HDA) ices: by compressing Ice~I up to 10 kbar at $\approx$ 80~K one obtains HDA instead of  
Ice~VI,\cite{mishima1984melting}
which may be transformed into LDA by decompression of HDA at 130 K~\cite{klotz2005t} or by heating recovered HDA at ambient pressure to beyond 130 K;
\cite{mishima1998relationship}
finally Ice~I is recovered by heating up LDA.
Similar connections between crystalline and amorphous ices are found in the high-pressure region of the water phase diagram where a very-high-density amorphous (VHDA) ice, plastic ices and crystalline structures with complex hydrogen-bond network (e.g. Ice VII) have been observed or
predicted.~\cite{amann2016colloquium,himoto2014yet} 

Molecular dynamics (MD), a simulation method that yields the atomic trajectories as a function of time at given thermodynamic conditions, is in principle able to track such transitions. The kinetic barriers are however generally too large to allow an efficient exploration of the configuration space within typical MD timescales. 
Hence, so far it has been necessary to introduce (i) simplistic interaction models~\cite{moore2011structural} and/or (ii) seeding techniques~\cite{espinosa2016time}. 
Another approach consists in using enhanced sampling techniques that accelerate the occurrence of rare events by focusing on low-dimensional order parameters, also called collective variables (CV) \cite{review}. 
Yet the CVs available to describe phase transitions are specifically designed for a given type of structural
transformation~\cite{lechner2008accurate,martovnak2003predicting,haji2015direct}, while no
general CV scheme has been proven successful for a wide class of problems, in particular those involving amorphous systems.
Recently, distance metrics developed for condensed
matter~\cite{valle2010crystal,gallet2013structural,pietrucci2015systematic,pietrucci2015formamide,zhu2016fingerprint,de2016comparing} 
have been proven to be successful in classifying structures in molecular or extended systems based on their atomic environment and/or interatomic network.
Here we show that by combining enhanced sampling techniques with a novel CV based on a general metric we are able to define in an efficient way the topological space of transformations among liquid, amorphous, and crystalline forms of water. This allows exploring at will the phase diagram along pathways connecting minimum-energy structures with a single, general approach, capable to characterize mechanisms and energetics. 

Our CV scheme relies on the concept of \textit{generalized distances} between configurations of the system, as defined from relative atomic positions, and it only requires to postulate the initial and final states of the target transformation, without any assumption on the pathway. The transformation is represented in a two-dimensional space $\{s,z\}$ of path CVs \cite{branduardi2007b}, with $s$ quantifying the progress of the transformation and $z$ allowing to discriminate between different pathways and to represent transitions to states that are not the target ones. 
In such a CV scheme each configuration of the system is associated with a permutation invariant vector \cite{gallet2013structural} (PIV), built-up from inter-atomic Cartesian distances. The PIV is built starting from atom-type-specific ordered blocks, $\v{v}_{kk'}$, with elements
\begin{align}\label{eq:PIV}
    v^{\beta\beta'}_{kk'}=c_{kk'}~\mathcal{S}\left(\sqrt[3]{\frac{\Omega_0}{\Omega}}\left|\v{r}_{\beta k}-\v{r}_{\beta' k'}\right|\right) ~ .
\end{align}
Here $\v{r}_{\beta k}$ is the position vector of the $\beta$-th atom of type $k$ (oxygen or hydrogen), with $\beta>\beta'$, $k>k'$; $c_{kk'}$ are coefficients that define the PIV variant (they are all equal to one in the original formulation); $\Omega$ and $\Omega_0$ are the volume of the simulation box and a reference volume (details below) respectively; $\mathcal{S}$ is a switching function monotonically decreasing from one to zero as $|\v{r}_{\beta k}-\v{r}_{\beta' k'}|$ increases (see Ref. \cite{suppInfo} for more details). 
In order to define the PIV, first the $v^{\beta\beta'}_{kk'}$ elements of each $\v{v}_{kk'}$ block are sorted in ascending ordered, then the different blocks are simply joined together resulting in a PIV of $N_\t{atoms}(N_\t{atoms}-1)/2$ components, that we indicate with $V_{\alpha}$. The sorting operation within each block introduces invariance upon permutation of identical atoms. The volume scaling factor (absent in Ref. \cite{gallet2013structural}) was found to be important to avoid violent fluctuations of the cell parameters during metadynamics. 

Distances between generic configurations X and Y are computed as squared Euclidean distances between the corresponding PIVs ($\mathcal{D}_\t{YX} =  \sum_{\alpha} (V_{\t{Y}\alpha} - V_{\t{X}\alpha})^2 $), and used to map each configuration of the system (X) into a point $(s_\t{X},z_\t{X})$ in the 2D space defined by the \textit{path} CVs \cite{branduardi2007b}, built starting 
from only two reference configurations A and B representing the initial and final state of the transformation:
\begin{align}
s_\t{X} &= \frac{1 \cdot e^{-\lambda\mathcal{D}_\t{AX}} + 2 \cdot e^{-\lambda\mathcal{D}_\t{BX}}} {e^{-\lambda\mathcal{D}_\t{AX}} + e^{-\lambda\mathcal{D}_\t{BX}}}\\
z_\t{X} &= -\lambda^{-1} \log \left( {e^{-\lambda\mathcal{D}_\t{AX}} + e^{-\lambda\mathcal{D}_\t{BX}}} \right) \ . 
\end{align}
The two coordinates track the progress from A to B and the distance from A and B, respectively. We remark that this formulation does not contain any guess about the mechanism of the transformation, and that the freedom granted by the $z$ coordinate allows to explore also the formation of unexpected metastable structures different from A and B (e.g., Ice VII-P in Figure 4-c). There is some freedom in the choice of the parameter $\lambda$: in this work we adopted $\lambda \simeq 2.3 / \mathcal{D}_\t{AB}$, conveniently localizing the free energy basins of reference states A and B around $s\simeq1.1$ and $s\simeq1.9$, respectively, and leading to smooth transformation pathways and landscapes. 
A much larger $\lambda$ would produce very irregular and discontinuous pathways, while a much smaller one would hamper the resolution of different phases.
We used metadynamics~\cite{laio2002escaping} for the discovery of continuum pathways between locally stable configurations, and umbrella 
sampling~\cite{torrie1977nonphysical} for the reconstruction of precise free energy landscapes. Computational details are given in Ref.~\cite{suppInfo}.

\begin{figure}
\centering\includegraphics[width=0.8\columnwidth]{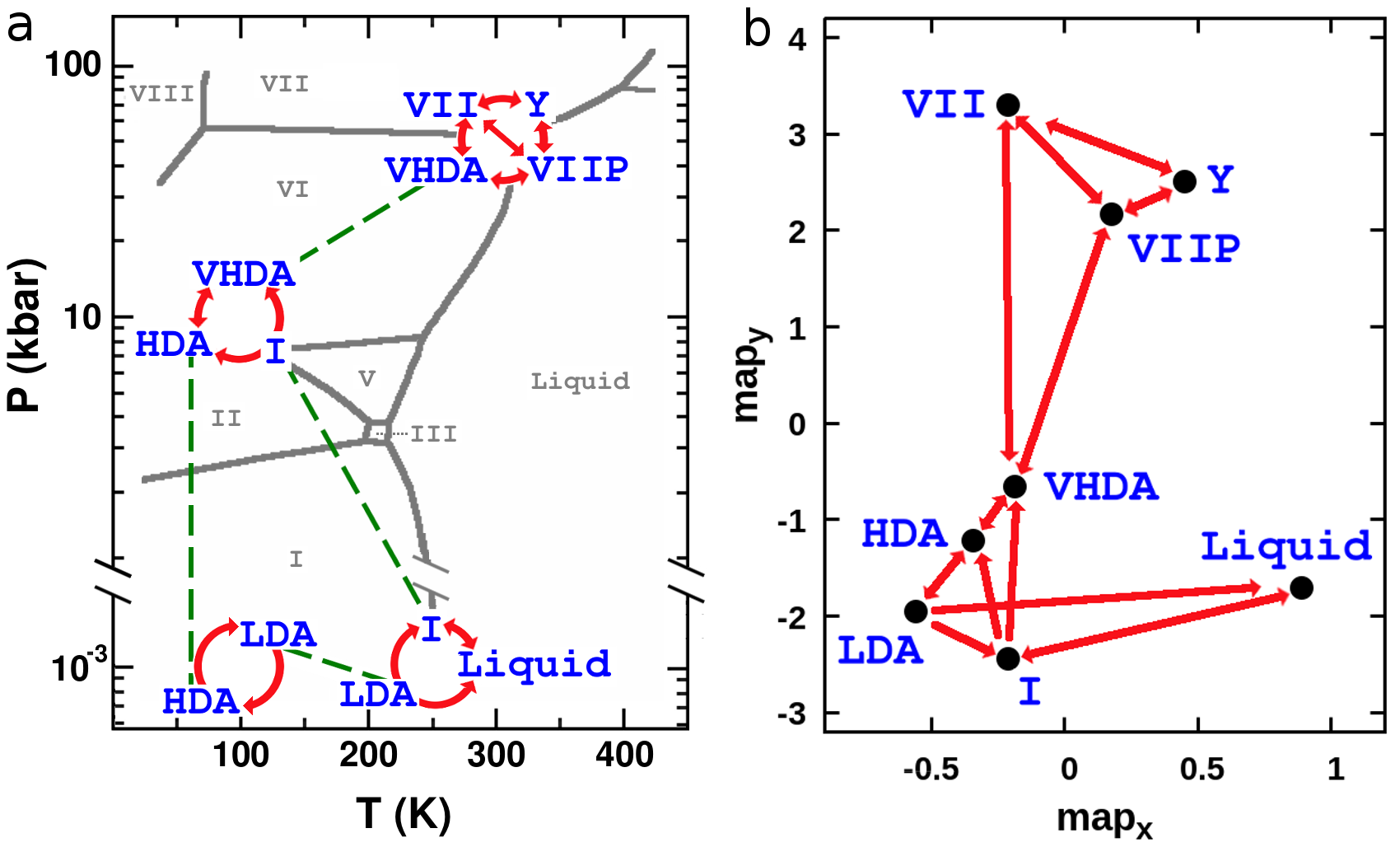}
\caption{\label{fig:path} 
\textbf{a}, The TIP4P/2005 water phase diagram \cite{abascal2005general} is shown in grey, and phase transitions between (meta)stable phases (blue labels) simulated with metadynamics are indicated with red arrows. Dashed green lines represent variations of the (P,T) conditions of the system within a phase, performed with unbiased molecular 
dynamics simulations. \textbf{b}, Two-dimensional map reproducing distances between PIV vectors defined in eq. (\ref{eq:PIV}). The map axes are defined within an arbitrary rotation.}
\end{figure}
In Figure \ref{fig:path}-a we draw the pathways we followed on the phase diagram of a realistic model of water \cite{abascal2005general}, \textit{navigating} within and across free-energy basins using standard MD and enhanced sampling techniques, respectively. Figure \ref{fig:path}-b shows a two-dimensional map (see also Ref.~\cite{pietrucci2015systematic}) of distances between the visited crystalline and amorphous structures: the metric employed to define the CVs is able to scatter the different phases in a way that recalls the topology of the phase diagram, representing kinetically connected phases as neighbors, and kinetically disconnected ones as far apart.

\begin{figure*}
\includegraphics[width=0.8\linewidth]{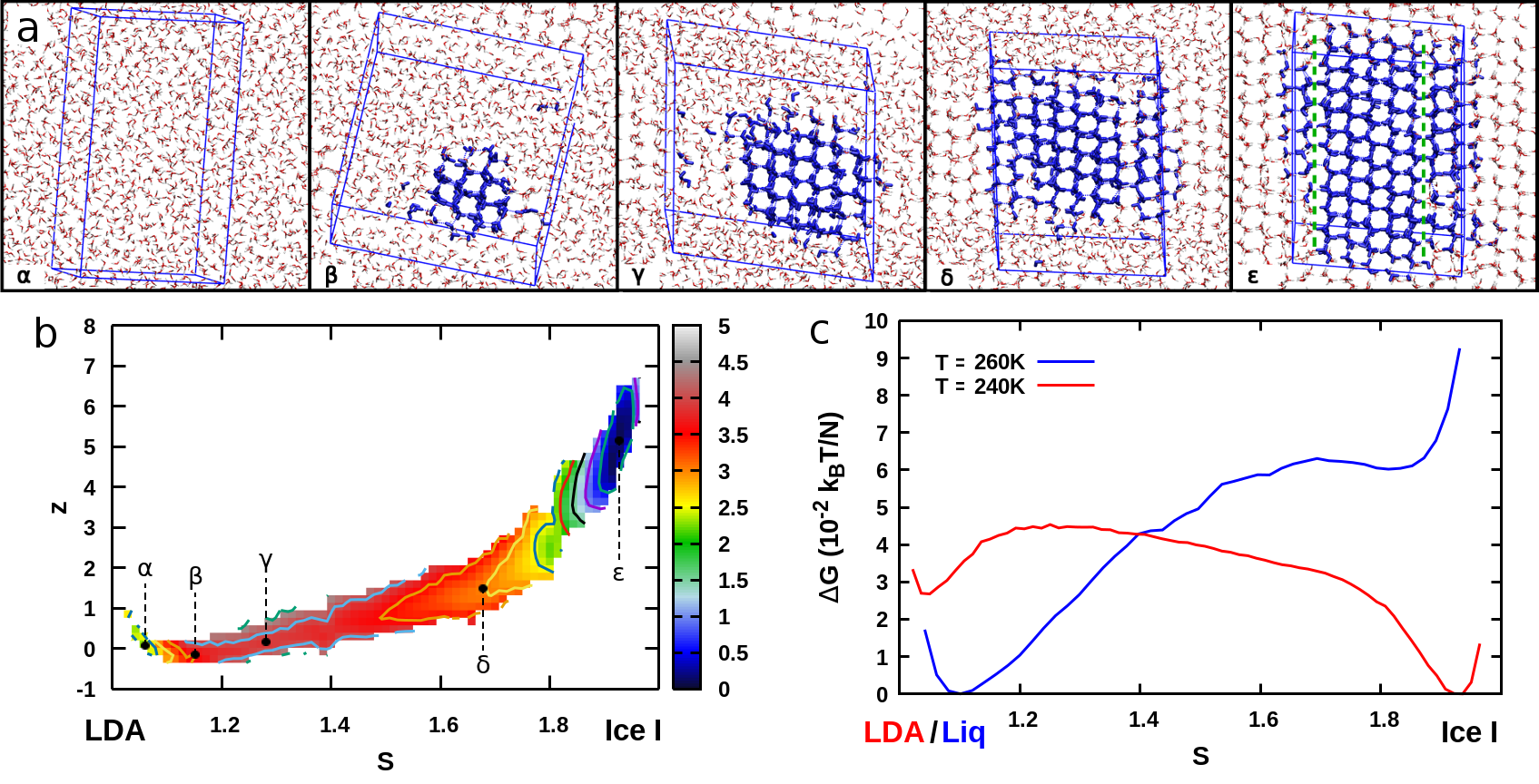}
\caption{\label{fig:cryst} 
\textbf{a}, Sequence of snapshots ($\alpha,\beta,\gamma,\delta,\epsilon$) of the umbrella sampling simulation describing the progressive crystallization of LDA water into Ice~I. Molecules in icelike environment, with an average local tetrahedral bond order parameter \cite{lechner2008accurate} of oxygen higher than 0.7, are shown in blue (in the box only). Snapshot $\gamma$ shows the Ice~I nucleus at its critical size (saddle point in the free-energy profile). Snapshot $\epsilon$ shows stacking disordered ice I, with Ic cubic regions separated by Ih hexagonal layers (green lines). \textbf{b}, Free-energy profile along the LDA-Ice~I transformation pathway ($T=240$~K,$P=1$~bar)  obtained via the weighted histogram analysis method applied to umbrella sampling trajectories. \textbf{c}, Comparison of the free-energy profiles, projected along the $s$ coordinate, for the crystallization transitions LDA -- Ice~I at $T=240$~K and $P=1$~bar and Liquid -- Ice~I at $T=260$~K and $P=1$~bar. The free energy minima of each simulation are arbitrarily set to zero ($N$ is the number of water molecules, here 800).}
\end{figure*} 

As a starting point we analyze the crystallization of Ice~I both from the liquid and the LDA phases at $P=1$~bar and over a range of temperatures around the melting point ($T_m\simeq250$~K for the adopted interatomic potential~\cite{abascal2005general}). 
The initial configurations have been respectively obtained by cooling down an equilibrium liquid phase from $T=300$~K and heating up a LDA structure from $T=100$~K \cite{suppInfo}.
Both crystallization transitions have been achieved multiple times in metadynamics simulations
at $T=240$~K and $T=260$~K and they are all characterized by (i) a nucleation mechanism that we show in Figure \ref{fig:cryst} for the LDA--Ice~I transformation at $T=240$~K, (ii) the formation of a crystal nucleus of cubic symmetry (Ice Ic), and (iii) a final state with either a perfect cubic symmetry or made up of layers of cubic Ic and hexagonal Ih ice (see last snapshot in Figure \ref{fig:cryst}-a). This last feature is in agreement with experimental findings~\cite{malkin2012structure}, and our results show that the formation of stacking disordered Ice~I may also proceed via the merging of Ic-nuclei as well as via (i) random growth of Ic and Ih layers~\cite{malkin2015stacking} and (ii) direct formation of Ic-Ih nuclei \cite{haji2015direct}. In Figure \ref{fig:cryst}-b we display the free energy profiles for the liquid--Ice~I and LDA--Ice~I transformations, respectively above and below the melting point. The calculated relative stabilities of the various phases agree with the phase diagram of the water model, and the order of magnitude of the free-energy profiles is consistent with free-energy differences calculated in previous works \cite{vega2008determination}. 
We remark that crystallizing the liquid above the melting temperature, in the bulk, without any seeds and with a very realistic water model, shows that our approach allows to perform very challenging transformations even in unfavorable conditions, reaching metastable states (here Ice I) starting from the global minimum. 

\begin{figure}
\includegraphics[width=0.7\columnwidth]{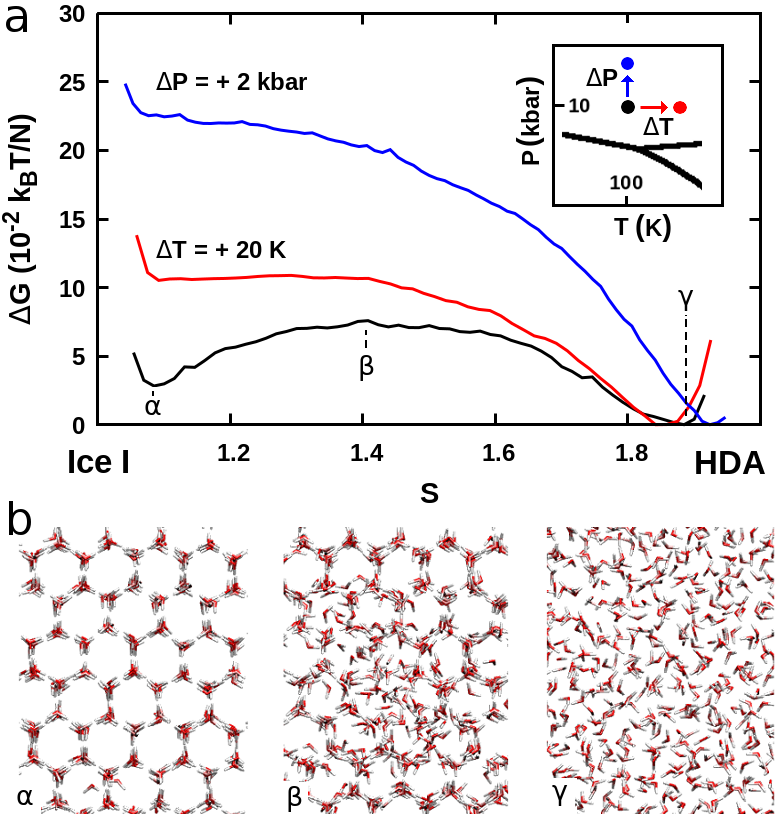}
\caption{\label{fig:amorphous} 
\textbf{a}, Free-energy profiles projected along the path collective variable $s$ for the Ice~I -- HDA transformation at different thermodynamic conditions: $T=100$~K and $P=10$~kbar (black), $T=120$~K and $P=10$~kbar (red), $T=100$~K and $P=12$~kbar (blue). The free energy of HDA is arbitrarily set to zero. The inset shows the position of the three transformations in the phase diagram: at $T=100$~K and $P=10$~kbar Ice~I is metastable (see Fig. 1-a) and separated from HDA by a barrier. \textbf{b}, Sequence of snapshots ($\alpha,\beta,\gamma$) along the amorphization process at $T=100$~K and $P=10$~kbar. They are taken from the umbrella sampling trajectories and their position along the $s$ coordinate is shown in panel a.}
\end{figure}

In order to carry on our \textit{continuous journey} towards high pressure, we follow the experimental routes by cooling down Ice~I to $T=100$~K, and then compressing it at $P=10$~kbar using standard MD: in the neighborhood of this point of the phase diagram we explore (with enhanced sampling simulations) the transformation to HDA as a function of temperature and pressure. We find that the free-energy barrier decreases when temperature and pressure increase, as we show in Figure \ref{fig:amorphous}. Furthermore we note that although the free energy of Ice~I is higher than the one of HDA already at $P=10$~kbar and $T=100$~K (which is expected since Ice~I is not the stable phase in this region of the phase diagram), the free-energy barrier is nonzero.
This result is in accordance with a common interpretation of the Ice~I--HDA transformation~\cite{mishima1998relationship,chen2011high} that can be seen as an extrapolation to low temperatures and high pressures of the Ice~I--Liquid coexistence line. Even if this description is mostly qualitative, it certainly invigorates the idea that low-density crystalline ice is unstable with respect to denser disordered forms at high pressures, and that such instability occurs at higher pressures as temperature is decreased.

\begin{figure*}
\includegraphics[width=0.8\linewidth]{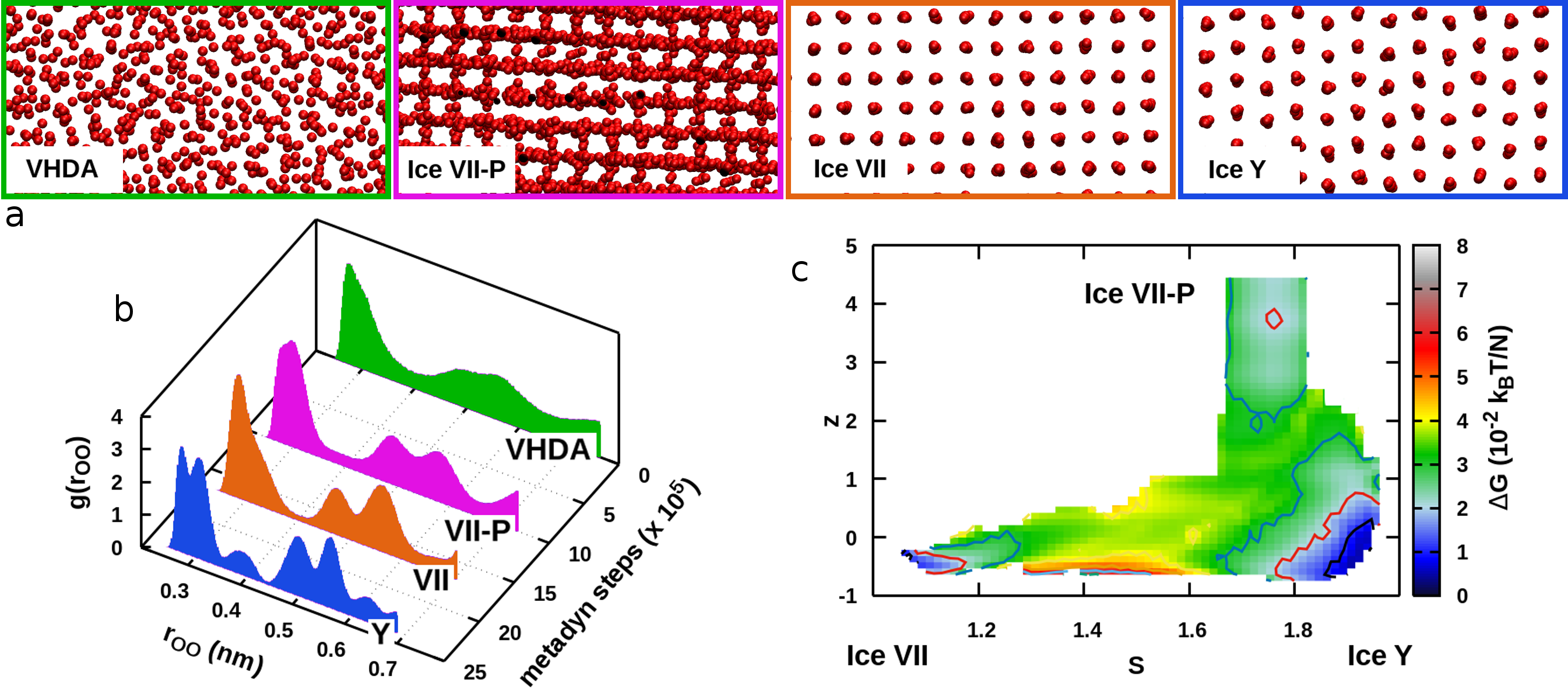}
\caption{\label{fig:HP} 
\textbf{a}, Snapshots representing the oxygen atoms of the system in four different configurations: VHDA, Ice~VII-P (plastic phase), Ice~VII and Ice~Y. All these structures are visited by the system while simulating the VHDA -- Ice~VII transformation at $T=300$~K and $P=50$~kbar via metadynamics. \textbf{b}, Oxygen--oxygen radial distribution functions as a function of the number of metadynamics simulation steps, for the stable configurations shown in panel a. The color code is consistent with panel a. \textbf{c}, Free energy landscape at $T=300$~K and $P=50$~kbar in the $\{s,z\}$ space obtained from umbrella sampling simulations. Equilibrated Ice~VII and Ice~Y are used to define the path collective variables. The simulation box includes $N=360$ water molecules.}
\end{figure*}

The HDA phase we obtain is then (i) decompressed at ambient pressure and transformed into LDA to close the \textit{loop} of transitions at low pressure, (ii) compressed at $P=12$~kbar and transformed into VHDA. 
Free-energy and density profiles for HDA-LDA and HDA-VHDA transformations are provided~\cite{suppInfo}. 
The HDA-VHDA transformation connects the low-pressure and high-pressure regions that we explore in this work.
The VHDA phase is compressed and heated until it reaches $P=50$~kbar and $T=300$~K, and at this point we address its crystallization to Ice~VII. While simulating the VHDA--Ice~VII transformation via metadynamics we observe that the system visits two additional metastable configurations. This result demonstrates that our method does not constrain the system to sample configurations along a simple path connecting the two reference structures, but rather allows it to follow complex mechanisms and discover new free energy basins. The metastable structures differ markedly from the other phases, as shown in Figure \ref{fig:HP} where we compare representative snapshots and the oxygen-oxygen radial distribution functions. The first metastable phase is identified as the plastic Ice~VII-P, which had already been proposed by Himoto {\it et al.}~\cite{himoto2014yet}. Oxygen atoms are arranged in a rather ordered crystalline network, whereas the hydrogen bond network changes dynamically: the correlation decay time of molecular dipoles is more than one order of magnitude shorter than that of Ice~VII at the same thermodynamic conditions. The second metastable phase (labeled here ``Ice~Y'') is characterized by a tetragonal oxygen lattice and stacked layers of hydrogen-bond networks. 
We leave a detailed investigation of this phase for future works (the atomic coordinates are provided~\cite{suppInfo}).
The flexibility of our method in representing transformations among several states in a two-dimensional CV space is illustrated in the free energy landscape connecting the three crystalline phases (Ice~VII, Ice~VII-P and Ice~Y) shown on Figure \ref{fig:HP}-c.

In conclusion, we propose a versatile method allowing the efficient simulation of phase transitions 
in condensed matter. 
We illustrated the approach by tackling the difficult problem of poly(a)morphism in water,
including kinetically challenging amorphous-to-crystalline and liquid-to-crystalline transitions. 
In particular, we simulated transformations between the LDA, HDA, liquid and Ice~I phases 
in the low-pressure region of the phase diagram and among VHDA and Ice~VII in the high-pressure region. 
In both cases, important mechanistic informations could be extracted from the simulations, 
highlighting the role of metastable structures during phase transitions.  
Thanks to a very general formulation, the proposed approach is not restricted to specific transitions 
of a single material.
Analysis of 50 experimental polymorphs belonging to 13 different materials 
(covalent, metallic, ionic, and molecular) 
indicates that the PIV-based metric is able to resolve all physically-distinct structures \cite{suppInfo},
suggesting a broad applicability of our simulation approach.
The ability to discover transformation mechanisms, simulate nucleation events, and reconstruct free energy landscapes and kinetic barriers -- all in a robust and systematic
way -- is highly complementary to structure prediction and materials discovery efforts.

\begin{acknowledgments}
FP thanks Gr\'egoire A. Gallet for help with the implementation of a preliminary version of the PIV path collective variables.
This work was funded within the {\it Investissements d'Avenir} program under reference ANR-11-IDEX-0004-02, within the framework of the cluster of excellence {\it MAT\'eriaux Interfaces Surfaces Environnement} (MATISSE) led by Sorbonne Universit\'es. We acknowledge calculations performed on the Gnome cluster at UPMC, on the Occigen cluster at CINES, Montpellier, France, and on the Ada cluster at IDRIS, Orsay, France, under GENCI allocations
2015-091387 and 2016-091387.
\end{acknowledgments}

\subfile{SI.tex}

\end{document}

%% file: SI.tex
\widetext

\author{S.~Pipolo}
\email{silvio.pipolo@univ-lille.fr}
\altaffiliation{Present affiliation: Univ. Lille, CNRS, Centrale Lille, ENSCL, Univ. Artois, UMR 8181 - UCCS - Unit\'e de Catalyse et Chimie du Solide, F-59000 Lille, France}
\affiliation{Sorbonne Universit\'es, UPMC Univ. Paris 06, CNRS UMR 7590, IRD UMR 206, MNHN, IMPMC, F-75005 Paris, France}
\author{M.~Salanne}
\affiliation{Sorbonne Universit\'es, UPMC Univ. Paris 06, CNRS, Laboratoire PHENIX, F-75005 Paris, France}
\author{G.~Ferlat}
\author{S.~Klotz}
\author{A.M.~Saitta}
\author{F.~Pietrucci}
\email{fabio.pietrucci@impmc.upmc.fr}
\affiliation{Sorbonne Universit\'es, UPMC Univ. Paris 06, CNRS UMR 7590, IRD UMR 206, MNHN, IMPMC, F-75005 Paris, France}

   \title{\LARGE Navigating at will on the water phase diagram}
   \author{S.~Pipolo}
   \email{silvio.pipolo@univ-lille1.fr}
   \altaffiliation{Now at: Unit\'e de Catalyse et Chimie du Solide, Universit\'e de Lille 1, 59655 Villeneuve d'Ascq, France}
   \affiliation{Sorbonne Universit\'es, UPMC Univ. Paris 06, CNRS UMR 7590, IRD UMR 206, MNHN, IMPMC, F-75005 Paris, France}
   \author{M.~Salanne} 
   \affiliation{
   Sorbonne Universit\'es, UPMC Univ. Paris 06, CNRS, Laboratoire PHENIX, F-75005 Paris, France}
   \author{G.~Ferlat} 
   \author{S.~Klotz} 
   \author{A.M.~Saitta}
   \author{F.~Pietrucci}
   \email{fabio.pietrucci@impmc.upmc.fr}
   \affiliation{Sorbonne Universit\'es, UPMC Univ. Paris 06, CNRS UMR 7590, IRD UMR 206, MNHN, IMPMC, F-75005 Paris, France}
\renewcommand{\thefigure}{S\arabic{figure}}
\renewcommand{\thesection}{S\arabic{section}}

 \maketitle



\setcounter{figure}{0}

\begin{center}
{\LARGE Supplemental Material}
\end{center}

\section*{Computational details}

In all enhanced sampling simulations the PIV is defined employing the following switching function of interatomic distances $r_{ij}$:
\begin{align}\label{eq:swf}
  \mathcal{S}(r_{ij})=\frac{1-(r_{ij}/r_0)^n}{1-(r_{ij}/r_0)^m}.
\end{align}
The decay range of the switching function is defined in order to maximize the distance between the two target 
phases using the initial volume of the box for each simulation as the reference volume $\Omega_0$.
The set of parameters $\{r_0,n,m\}$ for eq. (\ref{eq:swf}) used in simulations of Figures 2, 3 and 4 are respectively $\{0.7,4,12\}$, $\{0.4,6,20\}$ and $\{0.5,8,18\}$ ($r_0$ values are in nm). The first set of parameters is also used for the illustrative map in Figure 1-b, that is constructed, starting from random 2D positions of the representative points, by minimizing the difference between PIV distances and map distances with a Monte Carlo procedure, until obtaining errors smaller than 6\%.
$c_\t{OO}=1.0$ and $c_\t{HH}=0.2$ (eq.(1)) are used to construct the PIVs for the crystallization of Ice~I, in all other transformations $c_\t{HH}=0$ since oxygen-oxygen terms are sufficient to represent the transformation pathways.

The potential used to model interactions between water molecules is TIP4P/2005 \cite{abascal2005general}, with periodically-repeated boxes of $N=800$ molecules (except for the 2D umbrella sampling in Figure 4 where $N=360$). 
MD and enhanced sampling simulations were performed in the $NPT$ ensemble.
Time constants for pressure \cite{berendsen1984molecular} and temperature \cite{bussi2007canonical} coupling of $0.02$ ps and $0.2$ ps were respectively used. 
Due to the sizable computational cost of PIV construction, enhanced sampling simulations are on average 10 (50) times slower than standard MD for a PIV built-up with oxygen (oxygen and hydrogen) atoms. 

Metadynamics simulations, of typical duration between $10$ and $100$~ns, allowed to overcome kinetic barriers and quickly explore many different transformations, discovering mechanisms as well as unexpected ice phases.
For instance, we could repeatedly simulate the crystallization of liquid water or of LDA water at different conditions above and below the melting point.
Typical width, height, and deposition stride of the repulsive Gaussians are $\sigma_s = 0.02$, $\sigma_z = 0.4$, $h=1.0$~kJ/mol, $\tau = 2.0$~ps. 
To obtain free energy landscapes of high statistical precision we exploited umbrella sampling simulations, 
seeded from the structures explored with metadynamics.
Each umbrella sampling window has a typical duration of $20$~ns, with the last $5$~ns (corresponding typically to more than 1000 times the autocorrelation decay time of the variables) employed to reconstruct the free energy landscape using the weighted histogram analysis method~\cite{roux1995calculation}.
In the landscapes of Figures 2 and 4 we restricted the sampling to the relevant regions of the landscape, i.e., those containing the transition pathways.
Statistical errors on free energies, estimated by bootstrapping or by cutting each trajectory in 5 segments and taking standard deviations, are smaller than $10^{-3} k_BT/N$. 
Periodic boundary conditions are applied and the 3D particle-mesh-Ewald approach is used for electrostatics with a real-space cutoff of $0.6$ nm (the same cutoff is used for van der Waals interactions). 
GROMACS 5.1.2 \cite{berendsen1995gromacs} and a modified version of the PLUMED 2 plugin \cite{tribello2014plumed} (soon available from www.plumed.org or upon request) were employed to perform $NPT$ MD and enhanced sampling simulations.

\section*{PIV distances between polymorphs of different types of materials}\label{sec:maps}

In this section we analyze PIV distances in a set of 50 experimental polymorphs of ionic, molecular, metallic, and covalent solids, 
including elements, binary, ternary and organic compounds. 
The aim is to test whether our methodology -- based on a general formulation 
that is not designed specifically for water -- might be applied to several different classes of materials.
All crystal structures are taken from the Crystallography Open Database (www.crystallography.net). 
In Tables \ref{tab:1}-\ref{tab:13} we report PIV distances between polymorphs of iron, silicon, sodium, carbon, phosphorus, sulphur, 
SiC, SiO$_2$, RbCl, Fe$_2$O$_3$, MgSiO$_3$, benzene, and paracetamol.
In all cases, to avoid system-dependent fine tuning, 
we included all atoms in the PIV definition and 
we tested {\sl the same set} of three different switching functions of interatomic distances, with $n=6$, $m=12$ and $r_0 =$ 2.5, 3.5, or 4.5 {\AA}, respectively. 
In the tables, PIV distances are scaled by the square root of the number of atoms, according to the definition in Ref. \cite{pietrucci2015systematic},
to facilitate their comparison across different materials.
Polymorphs are presented in order of decreasing density.
To provide a reference allowing to appreciate the good separation between different polymorphs, 
we performed 100 ps-long MD simulations at 300 K of a single polymorph of iron (2000 atoms,
embedded atoms potential \cite{Mendelev03}), silicon (2304 atoms, Stillinger-Weber potential \cite{Stillinger85}), 
and benzene (108 molecules, OPLS-AA potential \cite{Jorgensen96}),
finding a maximum value of PIV distance between snapshots of each system 
(i.e., a thermal spread) equal to 0.016, 0.004, and 0.005, respectively (with $r_0=3.5$ {\AA}).
With similar MD simulations we obtained reference structures for liquid iron and silicon, at 2500 K,
also reported in Tables \ref{tab:1} and \ref{tab:2} for comparison.

It has been pointed out that, from a mathematical point of view, sets of points can be artificially built displaying
a same set of sorted distances \cite{Bartok13}.
Nevertheless, as shown in this work as well as in Ref. \cite{pietrucci2015systematic},
the PIV-based metric is able to resolve, in all the crystallographic cases considered so far, the structures corresponding to
physically-distinct forms of a material (e.g., different polymorphs or different amorphous forms), 
whereas it does not separate the structures corresponding to physically-equivalent realizations 
of a same form (e.g., independent configurations belonging to a liquid, or to a same amorphous form).
Not only the former, but also the latter is a very convenient feature of the present approach,
leading to a compact and well-defined free energy basin even for liquid phases and amorphous forms.
In the same spirit, invariance under permutation of identical atoms conveniently overlaps physically-equivalent 
structures differing only by the labelling of atoms.  

\begin{figure*}[h!]
{\large Fe} \hspace{8cm} {\large Si}\\
\includegraphics[width=0.49\linewidth]{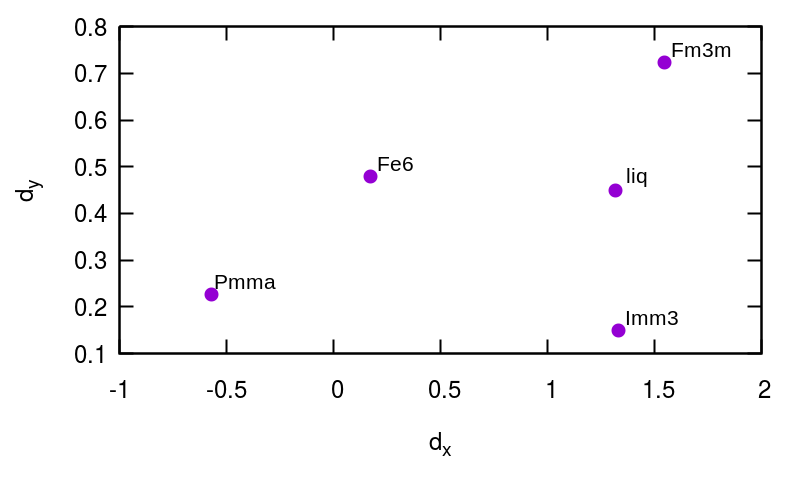}
\includegraphics[width=0.49\linewidth]{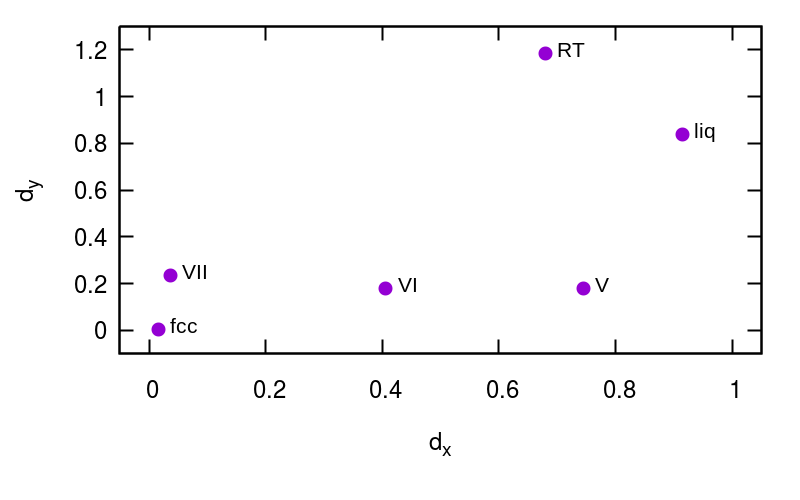}\\
{\large Fe$_2$O$_3$} \hspace{8cm} {\large MgSiO$_3$}\\
\includegraphics[width=0.49\linewidth]{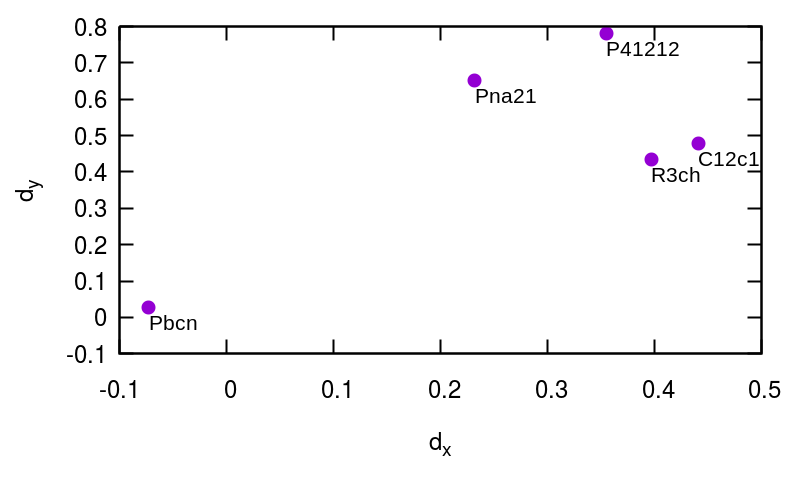}
\includegraphics[width=0.49\linewidth]{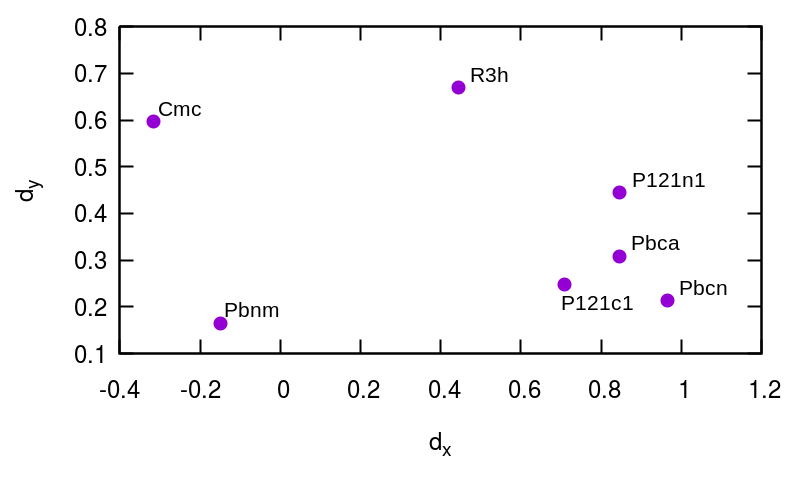}
\caption{\label{fig:PIVmaps} PIV distance 2D maps for Fe, Si, MgSiO$_3$ and Fe$_2$O$_3$ ($r_0=3.5$ {\AA}). 
Distances on the 2D map reproduce PIV distances, with an accuracy given by the following correlation coefficients:
0.9912 for Fe,  0.9905 for Si, 0.9964 for Fe$_2$O$_3$, and  0.9955 for MgSiO$_3$.}
\end{figure*}

In figure \ref{fig:PIVmaps} we show 2D maps of PIV distances \cite{pietrucci2015systematic}, generated using values in 
Tables \ref{tab:1}, \ref{tab:2}, \ref{tab:10}, and \ref{tab:11} with $r_0=3.5$ {\AA}. 
Note that the points size is larger than the maximum value of intra-polymorph PIV distances reported above.

\begin{table}[h!]
\centering
\caption{Iron PIV distances between different polymorphs, and a liquid model from MD.}
\label{tab:1}
\begin{tabular}{c | ccccc cccccc cccccc}
&\multicolumn{5}{c}{$r_0=2.5$ {\AA}} & & \multicolumn{5}{c}{$r_0=3.5$ {\AA}} & & \multicolumn{5}{c}{$r_0=4.5$ {\AA}}\\
\hline
\\
        & Fm3m & liq  &  Imm3 &Fe$_6$  & Pmma  & ~~~~  & Fm3m & liq  &  Imm3 &Fe$_6$  & Pmma& ~~~~ & Fm3m & liq  &  Imm3 &Fe$_6$  & Pmma \\
\hline
Fm3m    &   -  &  0.43 &  0.41 &  1.03 &  1.27 & ~~~~  &   -  &  0.41 &  0.58 &  1.43 &  2.12 & ~~~~  &   -  &  0.53 &  0.74 &  2.10 &  2.93 \\
liq     &  0.43 &   -  &  0.29 &  0.84 &  1.11 & ~~~~  &  0.41 &   -  &  0.34 &  1.18 &  1.90 & ~~~~  &  0.53 &   -  &  0.41 &  1.74 &  2.62 \\
Imm3    &  0.41 &  0.29 &   -  &  0.80 &  1.06 & ~~~~  &  0.58 &  0.34 &   -  &  1.18 &  1.92 & ~~~~  &  0.74 &  0.41 &   -  &  1.68 &  2.58 \\
Fe$_6$  &  1.03 &  0.84 &  0.80 &   -  &  0.28 & ~~~~  &  1.43 &  1.18 &  1.18 &   -  &  0.83 & ~~~~  &  2.10 &  1.74 &  1.68 &   -  &  1.15 \\
Pmma    &  1.27 &  1.11 &  1.06 &  0.28 &   -  & ~~~~  &  2.12 &  1.90 &  1.92 &  0.83 &   -  & ~~~~  &  2.93 &  2.62 &  2.58 &  1.15 &   -  \\
\end{tabular}
\end{table}

\begin{table}[h!]
\centering
\caption{Silicon PIV distances between different polymorphs, and a liquid model from MD.}
\label{tab:2}
\begin{tabular}{c | cccccc ccccccc ccccccc}
&\multicolumn{6}{c}{$r_0=2.5$ {\AA}} & & \multicolumn{6}{c}{$r_0=3.5$ {\AA}} & & \multicolumn{6}{c}{$r_0=4.5$ {\AA}}\\
\hline
\\
    & fcc & VII & VI & V   & RT  & liq &~~~~ & fcc & VII & VI & V   & RT  & liq &~~~~& fcc & VII & VI & V   & RT  & liq   \\
\hline
fcc  &  -  & 0.28 & 0.93 & 0.97 & 1.47 & 1.33 &~~~~ &  -  & 0.40 & 0.68 & 1.09 & 1.86 & 1.63 &~~~~ &  -  & 0.71 & 0.68 & 1.12 & 2.00 & 1.95 \\
VII  & 0.28 &  -  & 0.81 & 0.76 & 1.24 & 1.08 &~~~~ & 0.40 &  -  & 0.56 & 0.94 & 1.74 & 1.48 &~~~~ & 0.71 &  -  & 0.62 & 1.04 & 2.13 & 2.01 \\
VI   & 0.93 & 0.81 &  -  & 0.63 & 0.98 & 0.90 &~~~~ & 0.68 & 0.56 &  -  & 0.55 & 1.46 & 1.22 &~~~~ & 0.68 & 0.62 &  -  & 0.64 & 1.76 & 1.70 \\
V    & 0.97 & 0.76 & 0.63 &  -  & 0.87 & 0.66 &~~~~ & 1.09 & 0.94 & 0.55 &  -  & 1.34 & 1.06 &~~~~ & 1.12 & 1.04 & 0.64 &  -  & 1.44 & 1.34 \\
RT   & 1.47 & 1.24 & 0.98 & 0.87 &  -  & 0.42 &~~~~ & 1.86 & 1.74 & 1.46 & 1.34 &  -  & 0.66 &~~~~ & 2.00 & 2.13 & 1.76 & 1.44 &  -  & 0.61 \\
liq  & 1.33 & 1.08 & 0.90 & 0.66 & 0.42 &  -  &~~~~ & 1.63 & 1.48 & 1.22 & 1.06 & 0.66 &  -  &~~~~ & 1.95 & 2.01 & 1.70 & 1.34 & 0.61 &  -  \\
\end{tabular}                                                                                     
\end{table}

\begin{table}[h!]
\centering
\caption{Sodium PIV distances between different polymorphs.}
\label{tab:3}
\begin{tabular}{c | cc c cc c cc}
&\multicolumn{2}{c}{$r_0=2.5$ {\AA}} & & \multicolumn{2}{c}{$r_0=3.5$ {\AA}} & & \multicolumn{2}{c}{$r_0=4.5$ {\AA}}\\
\hline
\\
    & hcp  &  bcc   &~~~~   & hcp  &  bcc   &~~~~& hcp  &  bcc   \\
\hline
hcp &    -  &  0.07 & ~~~~  &    -  &  0.29 & ~~~~  &    -  &  0.45 \\
bcc &  0.07 &    -  & ~~~~  &  0.29 &    -  & ~~~~  &  0.45 &    -  \\
\end{tabular}                                                                                     
\end{table}

\begin{table}[h!]
\centering
\caption{Carbon PIV distances between different polymorphs.}
\label{tab:4}
\begin{tabular}{c | cccc c cccc c cccc}
&\multicolumn{4}{c}{$r_0=2.5$ {\AA}} & & \multicolumn{4}{c}{$r_0=3.5$ {\AA}} & & \multicolumn{4}{c}{$r_0=4.5$ {\AA}}\\
\hline
\\
&cubic&lonsdaleite&diamond&graphite& ~~~~  &cubic&lonsdaleite&diamond&graphite& ~~~~ &cubic&lonsdaleite&diamond&graphite  \\
\hline
cubic       &   -  &  2.83 &  2.87 &  3.34 & ~~~~  &   -  &  4.48 &  4.57 &  5.20 & ~~~~  &   -  &  5.54 &  5.63 &  6.40 \\
lonsdaleite &  2.83 &   -  &  0.22 &  0.82 & ~~~~  &  4.48 &   -  &  0.37 &  1.19 & ~~~~  &  5.54 &   -  &  0.38 &  1.22 \\
diamond     &  2.87 &  0.22 &   -  &  0.79 & ~~~~  &  4.57 &  0.37 &   -  &  1.13 & ~~~~  &  5.63 &  0.38 &   -  &  1.07 \\
graphite    &  3.34 &  0.82 &  0.79 &   -  & ~~~~  &  5.20 &  1.19 &  1.13 &   -  & ~~~~  &  6.40 &  1.22 &  1.07 &   -  \\
\end{tabular}
\end{table}

\begin{table}[h!]
\centering
\caption{Phosphorus PIV distances between different polymorphs.}
\label{tab-5}
\begin{tabular}{c | cccc c cccc c cccc}
&\multicolumn{4}{c}{$r_0=2.5$ {\AA}} & & \multicolumn{4}{c}{$r_0=3.5$ {\AA}} & & \multicolumn{4}{c}{$r_0=4.5$ {\AA}}\\
\hline
\\
&Pm3m&Cmca&Imma&P12c1& ~~~~  &Pm3m&Cmca&Imma&P12c1& ~~~~ &Pm3m&Cmca&Imma&P12c1  \\
\hline
Pm3m    &   -  &  0.32 &  0.50 &  0.52 & ~~~~  &   -  &  0.42 &  0.53 &  0.73 & ~~~~  &   -  &  0.38 &  0.59 &  0.94 \\
Cmca    &  0.32 &   -  &  0.50 &  0.53 & ~~~~  &  0.42 &   -  &  0.53 &  0.73 & ~~~~  &  0.38 &   -  &  0.51 &  0.90 \\
Imma    &  0.50 &  0.50 &   -  &  0.11 & ~~~~  &  0.53 &  0.53 &   -  &  0.33 & ~~~~  &  0.59 &  0.51 &   -  &  0.46 \\
P12c1   &  0.52 &  0.53 &  0.11 &   -  & ~~~~  &  0.73 &  0.73 &  0.33 &   -  & ~~~~  &  0.94 &  0.90 &  0.46 &   -  \\

\end{tabular}
\end{table}

\begin{table}[h!]
\centering
\caption{Sulfur PIV distances between different polymorphs.}
\label{tab:6}
\begin{tabular}{c | ccccc c ccccc c ccccc}
&\multicolumn{5}{c}{$r_0=2.5$ {\AA}} & & \multicolumn{5}{c}{$r_0=3.5$ {\AA}} & & \multicolumn{5}{c}{$r_0=4.5$ {\AA}}\\
\hline
\\
  & S ($\beta$) &  S7 (c) &  S7 (n)  & S12  & S8 &~~~~ & S ($\beta$) &  S7 (c) &  S7 (n)  & S12  & S8 &~~~~& S ($\beta$) &  S7 (c) &  S7 (n)  & S12  & S8 \\
\hline
S ($\beta$)&   -  &  1.20 &  1.21 &  1.23 &  1.25 & ~~~~  &   -  &  1.76 &  1.76 &  1.86 &  1.87 & ~~~~  &   -  &  2.50 &  2.51 &  2.60 &  2.61 \\
S7 (c)     &  1.20 &   -  &  0.02 &  0.05 &  0.06 & ~~~~  &  1.76 &   -  &  0.04 &  0.16 &  0.17 & ~~~~  &  2.50 &   -  &  0.06 &  0.19 &  0.19 \\
S7 (n)     &  1.21 &  0.02 &   -  &  0.06 &  0.06 & ~~~~  &  1.76 &  0.04 &   -  &  0.17 &  0.16 & ~~~~  &  2.51 &  0.06 &   -  &  0.19 &  0.19 \\
S12        &  1.23 &  0.05 &  0.06 &   -  &  0.03 & ~~~~  &  1.86 &  0.16 &  0.17 &   -  &  0.08 & ~~~~  &  2.60 &  0.19 &  0.19 &   -  &  0.10 \\
S8         &  1.25 &  0.06 &  0.06 &  0.03 &   -  & ~~~~  &  1.87 &  0.17 &  0.16 &  0.08 &   -  & ~~~~  &  2.61 &  0.19 &  0.19 &  0.10 &   -  \\
\end{tabular}                                                                                     
\end{table}    

\begin{table}[h!]
\centering
\caption{SiC PIV distances between different polymorphs (M stands for Moissanite).}
\label{tab:7}
\begin{tabular}{c | ccc c ccc c ccc}
&\multicolumn{3}{c}{$r_0=2.5$ {\AA}} & & \multicolumn{3}{c}{$r_0=3.5$ {\AA}} & & \multicolumn{3}{c}{$r_0=4.5$ {\AA}}\\
\hline
\\
       &M-3C & M-P6&  M-6H   & ~~~~  &M-3C & M-P6&  M-6H  & ~~~~ &M-3C & M-P6&  M-6H   \\
\hline
M-3C   &   -  &  0.10 &  0.22 & ~~~~  &   -  &  0.35 &  0.68 & ~~~~  &   -  &  0.57 &  0.92 \\
M-P6   &  0.10 &   -  &  0.21 & ~~~~  &  0.35 &   -  &  0.64 & ~~~~  &  0.57 &   -  &  1.05 \\
M-6H   &  0.22 &  0.21 &   -  & ~~~~  &  0.68 &  0.64 &   -  & ~~~~  &  0.92 &  1.05 &   -  \\

\end{tabular}
\end{table}

\begin{table}[h!]
\centering
\caption{SiO$_2$ PIV distances between different polymorphs.}
\label{tab:8}
\begin{tabular}{c | ccccc c ccccc c ccccc}
&\multicolumn{5}{c}{$r_0=2.5$ {\AA}} & & \multicolumn{5}{c}{$r_0=3.5$ {\AA}} & & \multicolumn{5}{c}{$r_0=4.5$ {\AA}}\\
\hline
\\
  &Coes.&Crist.&Quartz&Stishov.&Tridym.&~~~~ &Coes.&Crist.&Quartz&Stishov.&Tridym.&~~~~&Coes.&Crist.&Quartz&Stishov.&Tridym.\\
\hline
Stishov.   &   -  &  0.77 &  0.82 &  1.03 &  0.91 & ~~~~  &   -  &  1.03 &  1.09 &  1.41 &  1.49 & ~~~~  &   -  &  1.61 &  1.42 &  1.87 &  2.04 \\
Coes.      &  0.77 &   -  &  0.09 &  0.64 &  0.24 & ~~~~  &  1.03 &   -  &  0.25 &  0.57 &  0.61 & ~~~~  &  1.61 &   -  &  0.41 &  0.57 &  0.71 \\
Quartz     &  0.82 &  0.09 &   -  &  0.63 &  0.19 & ~~~~  &  1.09 &  0.25 &   -  &  0.50 &  0.51 & ~~~~  &  1.42 &  0.41 &   -  &  0.63 &  0.77 \\
Tridym.    &  1.03 &  0.64 &  0.63 &   -  &  0.59 & ~~~~  &  1.41 &  0.57 &  0.50 &   -  &  0.35 & ~~~~  &  1.87 &  0.57 &  0.63 &   -  &  0.31 \\
Crist.     &  0.91 &  0.24 &  0.19 &  0.59 &   -  & ~~~~  &  1.49 &  0.61 &  0.51 &  0.35 &   -  & ~~~~  &  2.04 &  0.71 &  0.77 &  0.31 &   -  \\
\end{tabular}                                                                                     
\end{table}    

\begin{table}[h!]
\centering
\caption{RbCl PIV distances between different polymorphs.}
\label{tab:9}
\begin{tabular}{c | cc c cc c cc}
&\multicolumn{2}{c}{$r_0=2.5$ {\AA}} & & \multicolumn{2}{c}{$r_0=3.5$ {\AA}} & & \multicolumn{2}{c}{$r_0=4.5$ {\AA}}\\
\hline
\\
    & Fm3m  &  Pm3m   &~~~~   & Fm3m  &  Pm3m &~~~~&  Fm3m  &  Pm3m   \\
\hline
Fm3m  &    -  &  0.23 & ~~~~  &    -  &  0.84 & ~~~~  &    -  &  1.19 \\
Pm3m  &  0.23 &    -  & ~~~~  &  0.84 &    -  & ~~~~  &  1.19 &    -  \\
\end{tabular}                                                                                     
\end{table}    

\begin{table}[h!]
\centering
\caption{Fe$_2$O$_3$ PIV distances between different polymorphs.}
\label{tab:10}
\begin{tabular}{c | ccccc c ccccc c ccccc}
&\multicolumn{5}{c}{$r_0=2.5$ {\AA}} & & \multicolumn{5}{c}{$r_0=3.5$ {\AA}} & & \multicolumn{5}{c}{$r_0=4.5$ {\AA}}\\
\hline
\\
  &Pbcn&R3ch&C12c1&Pna21&P41212&~~~~ &Pbcn&R3ch&C12c1&Pna21&P41212&~~~~&Pbcn&R3ch&C12c1&Pna21&P41212\\
\hline
Pbcn    &   -  &  0.39 &  0.41 &  0.53 &  0.58 & ~~~~  &   -  &  0.63 &  0.68 &  0.73 &  0.82 & ~~~~  &   -  &  1.05 &  1.20 &  1.10 &  1.37 \\
R3ch    &  0.39 &   -  &  0.04 &  0.29 &  0.35 & ~~~~  &  0.63 &   -  &  0.07 &  0.27 &  0.35 & ~~~~  &  1.05 &   -  &  0.20 &  0.27 &  0.44 \\
C12c1   &  0.41 &  0.04 &   -  &  0.27 &  0.33 & ~~~~  &  0.68 &  0.07 &   -  &  0.26 &  0.32 & ~~~~  &  1.20 &  0.20 &   -  &  0.31 &  0.36 \\
Pna21   &  0.53 &  0.29 &  0.27 &   -  &  0.15 & ~~~~  &  0.73 &  0.27 &  0.26 &   -  &  0.21 & ~~~~  &  1.10 &  0.27 &  0.31 &   -  &  0.36 \\
P41212  &  0.58 &  0.35 &  0.33 &  0.15 &   -  & ~~~~  &  0.82 &  0.35 &  0.32 &  0.21 &   -  & ~~~~  &  1.37 &  0.44 &  0.36 &  0.36 &   -  \\
\end{tabular}                                                                                     
\end{table}    

\begin{table}[h!]
\centering
\caption{MgSiO$_3$ PIV distances between different polymorphs.}
\label{tab:11}
\begin{tabular}{c | ccccccc c ccccccc c ccccccc}
&\multicolumn{7}{c}{$r_0=2.5$ {\AA}} & & \multicolumn{7}{c}{$r_0=3.5$ {\AA}}  \\
\hline
\\
    &Cmcm&Pbnm&R3h&P121c1&Pbca&P121n1&Pbcn&~~~~ &Cmcm&Pbnm&R3h&P121c1&Pbca&P121n1&Pbcn&~~~~&\\
\hline
Cmcm    &   -  &  0.32 &  0.60 &  0.87 &  0.91 &  0.89 &  0.97 & ~~~~  &   -  &  0.49 &  0.84 &  1.09 &  1.18 &  1.15 &  1.28   \\
Pbnm    &  0.32 &   -  &  0.50 &  0.76 &  0.79 &  0.77 &  0.85 & ~~~~  &  0.49 &   -  &  0.75 &  0.94 &  1.01 &  0.99 &  1.12   \\
R3h     &  0.60 &  0.50 &   -  &  0.50 &  0.52 &  0.53 &  0.57 & ~~~~  &  0.84 &  0.75 &   -  &  0.49 &  0.55 &  0.54 &  0.66   \\
P121c1  &  0.87 &  0.76 &  0.50 &   -  &  0.13 &  0.21 &  0.18 & ~~~~  &  1.09 &  0.94 &  0.49 &   -  &  0.20 &  0.24 &  0.29   \\
Pbca    &  0.91 &  0.79 &  0.52 &  0.13 &   -  &  0.14 &  0.16 & ~~~~  &  1.18 &  1.01 &  0.55 &  0.20 &   -  &  0.19 &  0.21   \\
P121n1  &  0.89 &  0.77 &  0.53 &  0.21 &  0.14 &   -  &  0.20 & ~~~~  &  1.15 &  0.99 &  0.54 &  0.24 &  0.19 &   -  &  0.26   \\
Pbcn    &  0.97 &  0.85 &  0.57 &  0.18 &  0.16 &  0.20 &   -  & ~~~~  &  1.28 &  1.12 &  0.66 &  0.29 &  0.21 &  0.26 &   -
\\
\\
& \multicolumn{7}{c}{$r_0=4.5$ {\AA}}\\
\hline
&Cmcm&Pbnm&R3h&P121c1&Pbca&P121n1&Pbcn\\
Cmcm    &   -  &  0.56 &  1.18 &  1.50 &  1.64 &  1.59 &  1.71 \\
Pbnm    &  0.56 &   -  &  1.05 &  1.33 &  1.45 &  1.42 &  1.54 \\
R3h     &  1.18 &  1.05 &   -  &  0.54 &  0.64 &  0.61 &  0.72 \\
P121c1  &  1.50 &  1.33 &  0.54 &   -  &  0.24 &  0.23 &  0.29 \\
Pbca    &  1.64 &  1.45 &  0.64 &  0.24 &   -  &  0.22 &  0.23 \\
P121n1  &  1.59 &  1.42 &  0.61 &  0.23 &  0.22 &   -  &  0.24 \\
Pbcn    &  1.71 &  1.54 &  0.72 &  0.29 &  0.23 &  0.24 &   -

\end{tabular}
\end{table}    

\begin{table}[h!]
\centering
\caption{Benzene PIV distances between different polymorphs.}
\label{tab:12}
\begin{tabular}{c | cc c cc c cc}
&\multicolumn{2}{c}{$r_0=2.5$ {\AA}} & & \multicolumn{2}{c}{$r_0=3.5$ {\AA}} & & \multicolumn{2}{c}{$r_0=4.5$ {\AA}}\\
\hline
\\
    &Pbca&P121c1&~~~~   &Pbca&P121c1&~~~~&Pbca&P121c1\\
\hline
Pbca  &    -  &  0.09 & ~~~~  &    -  &  0.17 & ~~~~  &    -  &  0.21 \\
P121c1  &  0.09 &    -  & ~~~~  &  0.17 &    -  & ~~~~  &  0.21 &    -  \\
\end{tabular}                                                                                     
\end{table}    

\begin{table}[h!]
\centering
\caption{Paracetamol PIV distances between different polymorphs.}
\label{tab:13}
\begin{tabular}{c | cc c cc c cc}
&\multicolumn{2}{c}{$r_0=2.5$ {\AA}} & & \multicolumn{2}{c}{$r_0=3.5$ {\AA}} & & \multicolumn{2}{c}{$r_0=4.5$ {\AA}}\\
\hline
\\
    &Pbca&P121n1&~~~~   &Pbca&P121n1&~~~~&Pbca&P121n1\\
\hline
Pbca    &    -  &  0.09 & ~~~~  &    -  &  0.23 & ~~~~  &    -  &  0.42 \\
P121c1  &  0.09 &    -  & ~~~~  &  0.23 &    -  & ~~~~  &  0.42 &    -  \\
\end{tabular}                                                                                     
\end{table}    
 

\pagebreak
\newpage
\section*{Amorphous--amorphous transformations}\label{sec:LDA-HDA}

\begin{figure*}[h!]
\hspace{-6.5cm}{\large\textbf{a}} \hspace{6.9cm} {\large\textbf{b}}\\
\includegraphics[width=0.4\linewidth]{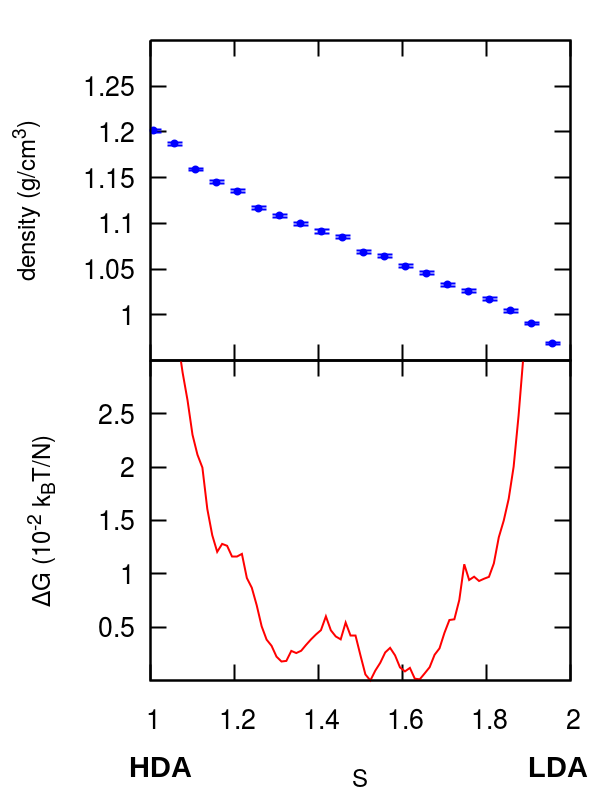}
\includegraphics[width=0.4\linewidth]{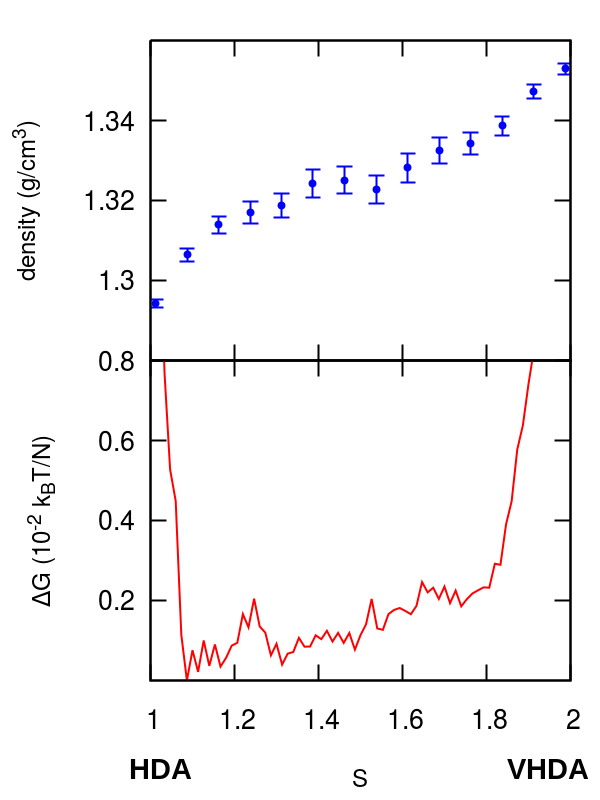}
\caption{\label{fig:am} \textbf{a}, Density (top) and free energy (bottom) profiles along the the HDA--LDA transformation pathway, represented by the path collective variable $s$, at $T = 100$~K and $P=1$~bar. $\Delta G$ values are obtained via the weighted histogram analysis method applied to the umbrella sampling trajectories (last $5$~ns of $20$~ns-long simulation windows); density values are averaged over umbrella sampling trajectories  (standard deviations are also reported). \textbf{b}, Same as panel a but for the HDA--VHDA transformation at $T = 100$~K and $P=12$~kbar.}
\end{figure*}

In this section we present an analysis of the HDA-LDA and HDA--VHDA transformations. In Figure \ref{fig:am} we report density and free energy profiles along the pathways of the two transformations obtained from umbrella sampling simulations. Density changes smoothly with the progress of the transformation, the profiles show several local minima but free energy barriers are typically smaller than the ones of the transformations reported in the manuscript, especially for the HDA-VHDA transformation, where they are of the order of magnitude of the estimated errors (see Methods section). This is in agreement with previous computational studies on the compression of the HDA form \cite{chiu2013pressure}, 
where no hysteresis behavior is observed during compression/decompression cycles. The set of parameters $\{r_0,n,m\}$ used to define the switching function for the HDA--LDA and HDA--VHDA simulations of Figure \ref{fig:am} are respectively $\{0.35,8,18\}$ and $\{0.40,6,20\}$. For more computational details see the Methods section.

\section*{Comparison of LDA and liquid: densities and diffusion coefficients}\label{sec:LDA}

We report in Table \ref{tab:LDA} the densities and the diffusion coefficients obtained from four 10 ns-long
MD simulations at P = 1 bar, of (i) LDA at T = 100 K (ii), LDA heated up at T = 240 K (iii), the liquid phase cooled down at T = 260 K, and (iv) the liquid phase at T = 300 K.

\begin{table}[h!]
\centering
\caption{Properties of liquid and LDA water at different conditions.}
\label{tab:LDA}
\begin{tabular}{c | cc }
& Density (kg/m$^3$) & Diffusion Coeff. ($10^{-5}$ cm$^2$/s) \\
\hline
LDA 100 K     & 987 $\pm$ 1 & 0.000041 $\pm$ 0.000008 \\
LDA 240 K     & 980.9 $\pm$ 0.3 & 0.29 $\pm$ 0.03 \\       
liq 260 K   & 985.2 $\pm$ 0.1 & 0.8407 $\pm$ 0.0009 \\
liq 300 K   & 976.3 $\pm$ 0.1 & 2.619 $\pm$ 0.008 \\
\end{tabular}                                                                                     
\end{table}


\newpage

%